\documentclass[osajnl,twocolumn,showpacs,10pt]{revtex4-1} 
\usepackage{amsmath,amssymb,graphicx}
\begin{document}

\title{Common mode noise rejection properties of amplitude and phase noise in a heterodyne interferometer}

\author{Gerald Hechenblaikner}\email{G.H.: Gerald.Hechenblaikner [at] astrium.eads.net}
\affiliation{EADS Astrium, 88039 Friedrichshafen, Germany}

\begin{abstract}
High precision metrology systems based on heterodyne interferometry can measure position and attitude of objects to accuracies of picometer and nanorad, respectively.
A frequently found feature of the general system design is the subtraction of a reference phase from the phase of the position interferometer, which suppresses low frequency common mode amplitude and phase fluctuations occurring in volatile optical path sections shared by both, the position and reference interferometer. Spectral components of the noise at frequencies around or higher than the heterodyne frequency, however, are generally transmitted into the measurement band and may limit the measurement accuracy. Detailed analytical calculations complemented with Monte Carlo simulations show that high frequency noise components may also be entirely suppressed, depending on the relative difference of measurement and reference phase, which may be exploited by corresponding design provisions. Whilst these results are applicable to any heterodyne interferometer with certain design characteristics, specific calculations and related discussions are given for the example of the optical metrology system of the LISA Pathfinder mission to space.
\end{abstract}


\maketitle


\section{Introduction}
\label{sec:intro}

Metrology systems based on heterodyne interferometry have numerous applications in different areas of measurement technology where high precision coupled with excellent stability and robust operation are required. For example, they have recently been used in precise angle measurements and characterization of beam alignment \cite{Hah2010},\cite{Mul2005}, measurement of thermal deformation and expansion (dilatometry) \cite{Niw2009},\cite{Cor2009},\cite{Kim2010}, in stellar interferometers for astrometric measurements \cite{Halv2002},\cite{Gou2004},\cite{Sha2009}, and in gravitational wave detectors \cite{Hei2003},\cite{Hec2011},\cite{Aud2011}.
Heterodyne interferometry (dual frequency) offers several distinct advantages over homodyne (single frequency) interferometry: The phase is read out from an AC rather than a DC interference signal, which facilitates and simplifies the detection and read-out process on the one hand, as well as avoiding detrimental exposure to low frequency noise on the other hand.
 The usage of Acousto-Optic-Modulators (AOMs) in heterodyne frequency generation\cite{Tan1989} has spurred a period of continuous improvements in measurement precision over the past decades \cite{Zha1999},\cite{Law2000},\cite{Wu2002}, reaching the picometer level for position and the nanorad level for angular measurements, respectively\cite{Cor2009},\cite{Aud2011},\cite{Hec2011}.

Accuracy and stability requirements for space-borne gravitational wave detectors are exceptionally demanding so that noise sources and performance limiting factors in the respective interferometers must be well characterized and appropriate mitigating measures taken. Whilst many discussions in this paper apply to heterodyne interferometers in general, specific numbers and design references are given for the example of the Optical Metrology System (OMS) of the LISA Pathfinder (LPF) space mission\cite{Vit2003}. LPF is a technological precursor to the Laser-Interferometer Space Antenna (LISA) mission which aims to detect gravitational waves with interferometry\cite{Bel2008} and the OMS is the most precise metrology systems qualified for space as of today\cite{Hei2003}\cite{Hec2011}.
As a fundamental feature of its design, the phase of a reference interferometer is subtracted from the phase of the actual measurement interferometer so that common mode noise fluctuations, which arise from unstable components and optical path sections  shared by both interferometers, are suppressed to a large extent. However, this only holds for low frequency fluctuations, whereas high frequency noise directly couples into the measurement band. Recent experimental tests observed a distinct variation of measurement performance depending on the relative phase between the two interferometer signals. The calculations and simulations performed in this paper demonstrate that the rejection of high frequency noise depends on the relative interferometer phase, which could explain the experimental data. Simple analytical expressions are found for amplitude and phase noise rejection which are in excellent agreement with the results of Monte Carlo simulations.

\section{Digital Heterodyne Interferometry}
In this section we give a short introduction to digital heterodyne interferometry and define concepts and parameters relevant to the discussions of subsequent sections.
\subsection{Interferometer Schematic}
\label{sec:schematic}
\begin{figure}
   \centerline{\includegraphics[width=0.8\columnwidth]{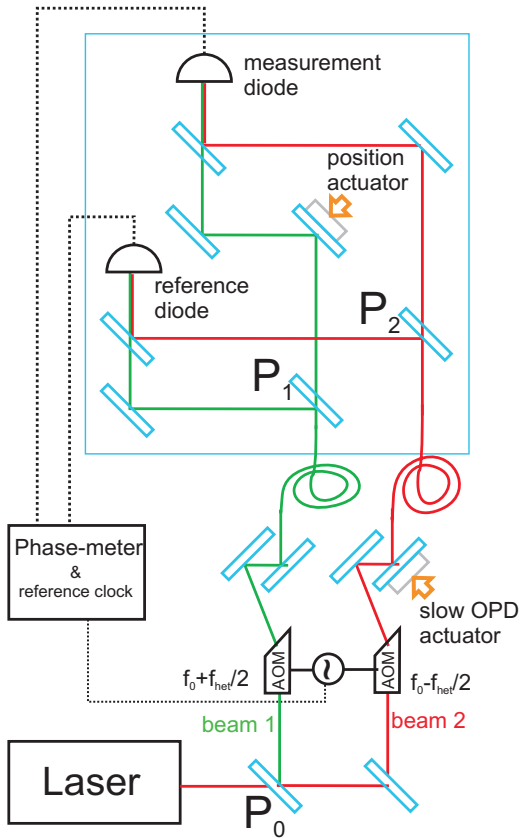}}
   \caption
    {(Color online) Schematic of a high-precision heterodyne interferometry scheme consisting of a measurement and reference interferometer. The laser output is split into beam 1, denoted by the green line on the left, and beam 2, denoted by the red line on the right, at beam splitter $P_0$. A common clock signal is shared between the heterodyne source and the phase-meter. \label{fig:opt_bench}}
\end{figure}

The basic design schematic (greatly simplified) of a high-precision optical heterodyne interferometer, similar to the one used in the LISA Pathinder (LPF) mission \cite{Hei2003}, is given in Fig.\ref{fig:opt_bench}. In order to facilitate the following discussion, the values for parameters applying to the optical metrology system of LPF are given in brackets as a baseline.\\
The output of a stable laser source is split into beam 1 and beam 2 at beam splitter $P_0$ before the 2 beams are shifted by the heterodyne frequency $f_{\rm het}$ (1 kHz) relative to another when passing through the respective Acousto-Optic-Modulators (AOMs).
Beams 1 and 2 are then coupled through optical fibres onto an ultra-stable optical bench where each beam is split again into separate components at beam splitters P1 and P2, respectively. The beam components reflected towards the left form the "reference interferometer" and are combined to interfere on the corresponding diode. Those components continuing along a straight line after the beam splitters are combined to interfere in the "position interferometer", appropriately named with respect to the variable phase which it aims to measure (represented by a position actuator). In a practical system this phase change could arise due to the expansion of a temperature sensitive component or the scan of an uneven surface \cite{Zha1999}. Alternatively, it could originate from the longitudinal movement of a reflective test-mass as in the basic operating scheme of the LPF interferometer\cite{Hei2003,Hec2011}.
Due to the frequency difference between beam 1 and 2 the interferometer signals $S_1(t)$ and $S_2(t)$ display an oscillation at the heterodyne frequency $f_{het}$ with a different phase offset $\phi$ and similar amplitude $A_{\rm in}$ for each interferometer.
\begin{equation}
S_1(t)=A_{\rm in}\left[1+\cos\left(2\pi f_{\rm het}t-\phi\right)\right]
\end{equation}
The interferometer signals are detected by photo-diodes and processed by a digital phase-meter which samples the signals at $f_{\rm samp}$ (50 kHz) and performs a Discrete Fourier Transform (DFT) for a single frequency bin corresponding to the heterodyne frequency $f_{\rm het}$. To this end, the phase-meter and the heterodyne signal source mixing into the AOM must share a common clock to which they are synchronized. Phase-meter data, consisting of the real and imaginary part of a complex phase vector $F_t$ are output after each DFT period $T$, corresponding to an update frequency $f_{\rm up}=1/T$ (100 Hz). The digital processing steps of the phase-meter to extract the signal phase are described in Eq.\ref{eqn:phase_int}:
\begin{eqnarray}
\Re(F_t)=\sum_{k=0}^{N-1} S(k\cdot\Delta t) \cos\left(\frac{2\pi m k }{N}\right)\nonumber\\
\Im(F_t)=\sum_{k=0}^{N-1} S(k\cdot\Delta t) \sin\left(\frac{2\pi m k }{N}\right),\label{eqn:phase_int}
\end{eqnarray}
where $N=$500 is the number of points in the DFT and the heterodyne frequency corresponds to bin number $m$ ($f_{het}=m~2\pi/T$). The vector components are then transmitted to a computer for further processing (e.g. calculation of phase and position or angular alignment). Note that the analyses and expressions for common mode noise rejection derived in the succeeding sections are based on the phase read-out scheme described by Eq.\ref{eqn:phase_int} which is found in many digital phase-meters of heterodyne interferometers.

The computer may also implement digital controllers for laser frequency, laser power and Optical Path-length Difference (OPD) stabilization \cite{Hec2011}, which will not be discussed further in this paper.
OPD stabilization was found to be necessary in the optical metrology system of LPF to mitigate the effects of optical sidebands arising from the cross-coupling of RF-signals going to the AOMs \cite{Wan2006}. The corresponding actuator for the servo-loop is depicted right after the AOM at the bottom of Fig.\ref{fig:opt_bench}.
\subsection{Noise Rejection Characteristics}
In order to obtain a highly stable position measurement, the phase $\psi_n$ of the reference interferometer is subtracted from the phase $\phi_n$ of the position interferometer, both of which are assumed to be affected by the same phase and amplitude noise fluctuations $n$.
Through this processing step the noise contributions of optical path sections shared by both interferometers are removed entirely from the difference phase $\phi_n-\psi_n$. The corresponding path sections in our schematic of Fig.\ref{fig:opt_bench} are those between the original beam splitter $P_0$ and the subsequent beam splitters $P_1$ and $P_2$ for beam 1 and beam 2, respectively, including the optical fibres and AOMs. These components are known to be quite unstable and sensitive to environmental changes such as temperature fluctuations so that a removal of the associated phase and amplitude fluctuations is critical for high performance.
Phase fluctuations occurring in the other path sections which are specific to each interferometer may be neglected as these sections are located on an ultra-stable optical bench (made from Zerodur) with minimal sensitivity to thermal expansion. The bench only comprises optical elements made from fused silica which are hydroxide-catalysis bonded\cite{Ell2005} to the Zerodur baseplate so that a quasi-monolithic structure with superior stability and negligible sensitivity to thermal expansion is formed.
However, such a noise cancellation scheme only applies to fluctuations at low frequencies (such as those induced by temperature swings) well below the phase-meter output frequency $f_{\rm up}$ (100 Hz). The question arises to what extent high frequency amplitude and phase noise (as often seen in optical fibres) couples into the measurement band and affects measurement performance. This will be discussed in detail for the remainder of this paper.

\section{Amplitude Noise}
\begin{figure}
   \centerline{\includegraphics[width=.8\columnwidth]{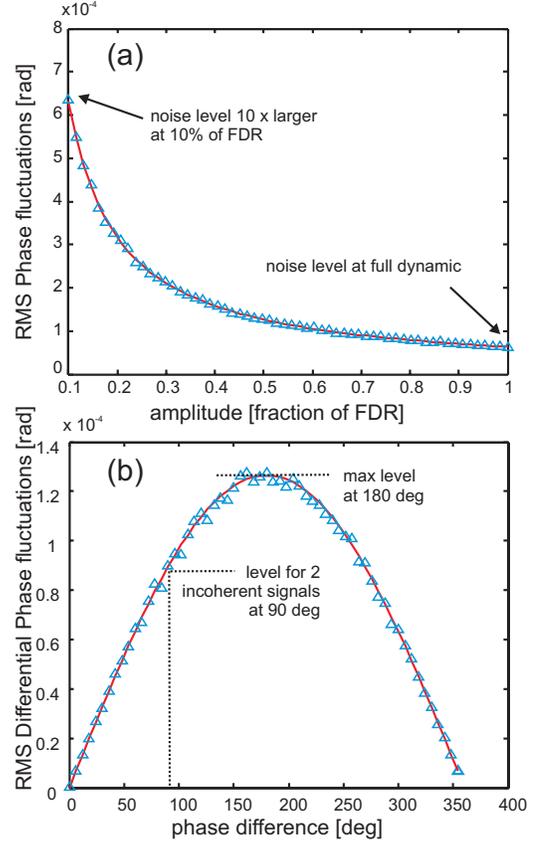}}
   \caption
    { \label{fig:amp_noise_plot}
(Color online) Impact of amplitude noise on output phase fluctuations:
(a) Output phase noise is plotted against input signal amplitude, given as a fraction of the full dynamic range (FDR), according to Eq.\ref{equ:AN_result} (red solid line). The results of Monte Carlo simulations are given by the open triangles. (b) Differential phase noise $\Delta n_{\rm an}$ is plotted against the mean phase difference $\phi-\psi$ between signal 1 and signal 2. RMS values of the amplitude noise at each sampling step are $10^{-3}{\rm rad}$ relative to the input amplitude. The predicted curve (Eq.\ref{equ:AN_key_equation}) is given by the red solid line, the results from Monte Carlo simulations are given by the blue triangles.}
\end{figure}

In the following we shall assume that a sinusoidal input signal is affected by amplitude noise with uniform linear spectral density $n_{\rm ld}~[{\rm V}/\sqrt{\rm Hz}]$. The corresponding fluctuations for each sampling step represent distinct random variables which are uncorrelated with each other and are assumed to have the following basic properties:
\begin{eqnarray}
\langle n_k\rangle&=&0\nonumber\\
\sqrt{\langle n_k n_m\rangle}&=&\delta_{km}n_{\rm ld}\sqrt{\frac{f_{\rm samp}}{2}}.\label{equ:noise_operators}
\end{eqnarray}
We shall determine the impact of the input amplitude noise on the phase error at the phase-meter output after the signal has been digitized and processed. Quantization noise shall be neglected in the derivations but is considered in the accompanying Monte Carlo simulations (at the level of the sensor noise described at the end of section 2).
To this end, we make the following ansatz for the real part (and an analogous one for the imaginary part) of the complex phase vector:
\begin{eqnarray*}
&&\Re(F_t)=\sum_{k=0}^{N-1} S(k\cdot\Delta t) \cos\left(\frac{2\pi m k }{N}\right)\nonumber\\
&&=\cos\left(\frac{2\pi m k }{N}\right)\sum_{k=0}^{N-1}\left[A_{\rm in} \cos\left(\frac{2\pi m k }{N}-\phi\right)+n_k\right]\nonumber\\
\end{eqnarray*}
where $n_k$ denotes the fluctuations of the amplitude noise (an) at a given sampling step $k$ and $A_{\rm in}$ is the amplitude of the input signal.
Applying some trigonometric simplifications and orthogonality relations we obtain after some algebra for the measured signal phase $\phi_{\rm an}$:
\begin{equation}
\tan\phi_{an}=\frac{\Im(F_t)}{\Re(F_t)}=\frac{\sin\phi+\frac{2}{NA_{\rm in}}\sum_{k=0}^{N-1}n_k\sin\left(\frac{2\pi m k }{N}\right)}{\cos\phi+\frac{2}{NA_{\rm in}}\sum_{k=0}^{N-1}n_k\cos\left(\frac{2\pi m k }{N}\right)}\label{equ:amp_noise_impact}
\end{equation}
It is now useful to rewrite Equation \ref{equ:amp_noise_impact} in a way that allows to extract the mean phase $\phi$ from the expression:
\begin{eqnarray}
\phi_{\rm an}&=&{\rm arg}\left\{e^{i\phi}+\sum_{k=0}^{N-1}\frac{2n_k}{N A_{\rm in}}e^{i2\pi m k/N}\right\}\nonumber\\
&=&\phi+
{\rm arg}\left\{1+\sum_{k=0}^{N-1}\frac{2n_k}{N A_{\rm in}}e^{i2\pi m k/N-\phi}\right\}.\label{equ:phase_extracted}
\end{eqnarray}
The noise fluctuations appearing in the sum of Eq.\ref{equ:phase_extracted} are tiny compared to 1 ($n_k\ll 1$) so that the argument of the complex number in brackets is given by the imaginary components alone. We therefore find for the output phase-fluctuations
\begin{equation}
\Delta\phi_{\rm an}=\phi_{\rm an}-\phi\approx\frac{2}{NA_{\rm in}}\sum_{k=0}^{N-1}n_k\sin\left(\frac{2\pi m k }{N}-\phi\right).\label{equ:amp_noise}
\end{equation}
and for the mean-square fluctuations
\begin{eqnarray}
\langle \Delta\phi_{\rm an}^2 \rangle&=&\frac{4 \langle n_k^2\rangle}{N^2A_{\rm in}^2}\sum_{k=0}^{N-1} sin^2\left(\frac{2\pi m k }{N}-\phi\right)\nonumber\\
&=&\frac{2\langle n_k^2\rangle}{N A_{\rm in}^2}.
\label{equ:amp_flucts}
\end{eqnarray}
Considering that the number of points N in the FFT is given by N=$f_{\rm samp}/f_{\rm up}$ we obtain for the phase-measurement noise induced by amplitude fluctuations
\begin{equation}
\sqrt{\langle \Delta\phi_{\rm an}^2 \rangle}=\sqrt{\frac{2\langle n_k^2\rangle}{N A_{\rm in}^2}}=\frac{n_{\rm ld}\sqrt{f_{\rm up}}}{A_{\rm in}}\label{equ:AN_result}
\end{equation}
and normalizing by the measurement band ($f_{up}/2$) we obtain for the uniform linear spectral density of the phase noise
\begin{equation}
{\rm LSD}(\Delta\phi_{\rm an})=\frac{n_{\rm ld}\sqrt{2}}{A_{\rm in}}
\end{equation}
As an example, in the optical metrology system of the LISA Pathfinder mission the uniform linear spectral density of the amplitude noise is given (with power stabilization loops on) by $n_{\rm ld}/A_{\rm in}=10^{-6}~{\rm Hz}^{-1/2}$ so that the resulting phase noise is around $1~\mu{\rm rad}/\sqrt{\rm Hz}$. This is smaller than ADC internal noise which is on the order of $10~\mu{\rm rad}/\sqrt{\rm Hz}$ for 12\% of  dynamic range \cite{Hec2011}, but may become appreciable for small signal amplitudes.
In Fig.\ref{fig:amp_noise_plot}a the output phase noise according to Eq.\ref{equ:AN_result} is plotted against signal amplitude given as a fraction of the Full Dynamic Range (FDR) for input amplitude noise of a fixed magnitude. The red solid line denotes the predictions of Eq.\ref{equ:AN_result} whereas the open triangles give the results of Monte Carlo simulations, where the standard deviation was found for 2000 simulation runs per amplitude value. The excellent agreement between the analytical formula and the simulations confirms the validity of Eq.\ref{equ:AN_result}.

\subsection{Common Mode Amplitude Noise Rejection}
We shall now investigate to what extent common mode amplitude fluctuations can be rejected between two signals depending on the relative phase between them and revert to Eq.\ref{equ:amp_noise} for that purpose. Assuming that the second signal has a phase $\psi$ with respect to the first signal and, without loss of generality, setting the relative phase of the first signal with respect to the phase of the phase-meter clock signal to zero, we obtain for the difference in fluctuations of the output phase
\begin{eqnarray}
&&\Delta n_{\rm an}=\Delta\phi_{\rm an}-\Delta\psi_{\rm an}\nonumber\\
&&=\sum_{k=0}^{N-1} \frac{2 n_k}{NA_{\rm in}}\left[\sin\left(\frac{2\pi m k}{N}\right)-\sin\left(\frac{2\pi m k}{N}-\psi\right)\right]\nonumber\\
&&=\sum_{k=0}^{N-1} \frac{2 n_k}{NA_{\rm in}}\left[\sin\left(\frac{2\pi m k}{N}\right)\left(1-\cos\psi\right)+\cos\left(\frac{2\pi m k}{N}\right)\sin\psi\right]\nonumber\\
\label{equ:AN_fundamental}
\end{eqnarray}
For the mean-square fluctuations of $\Delta n_{\rm an}$ we then obtain from Eq.\ref{equ:AN_fundamental}
\begin{eqnarray}
\langle \Delta n_{\rm an}^2\rangle &=&\frac{4}{N^2A_{\rm in}^2}\sum_{k=0}^{N-1}\langle n_k^2\rangle\sin^2\left(\frac{2\pi m k}{N}\right)\left(1-\cos\psi\right)^2\nonumber\\
&+& \frac{4}{N^2A_{\rm in}^2}\sum_{k=0}^{N-1}\langle n_k^2\rangle\cos^2\left(\frac{2\pi m k}{N}\right)\sin^2\psi\nonumber\\
&=&\frac{2}{NA_{\rm in}^2}\langle n_k^2\rangle 2\left(1-\cos\psi\right)=\frac{2}{NA_{\rm in}^2}\langle n_k^2\rangle 4\sin^2\left(\psi/2\right)\nonumber\\
\label{equ:AN_main_result}
\end{eqnarray}
Without loss of generality we can replace $\psi$ by $\phi-\psi$ in Eq.\ref{equ:AN_main_result}. Inserting the previous definition of the uniform linear spectral density of input noise into this equation we then obtain for the RMS fluctuations of the differential phase between signal 1 and signal 2
\begin{equation}
\sqrt{\langle \Delta n_{\rm an}^2\rangle}=\frac{n_{\rm ld}\sqrt{f_{\rm up}}}{A_{\rm in}}2\left|\sin\left(\frac{\phi-\psi}{2}\right)\right|
\label{equ:AN_key_equation}
\end{equation}
Comparing with Eq.\ref{equ:AN_result} we find that the output phase noise of the signal difference is increased by a factor of $2\sin[(\phi-\psi)/2]$ with respect to the output phase fluctuations of a single signal. This relationship is depicted in Fig.\ref{fig:amp_noise_plot}b, where the red solid curve is described by the analytical formula of Eq.\ref{equ:AN_key_equation} and the blue open triangles correspond to the results of Monte Carlo simulations. For these simulations the difference angle was varied by 360 degrees in steps of 6 degrees, whilst for each angle the standard deviation of the measured phase differences was found from 2000 simulation runs. \\
We find excellent agreement between the two methods. Depending on the relative signal phase, the noise in the phase difference may increase or decrease, reaching a minimum for both signals in phase and a maximum when the two signals are out of phase by 180 deg, as may intuitively be expected. When the 2 signals are 90 deg out of phase, the noise level for the difference signal reaches a value of $\sqrt{2}$ times the one for a single signal. Note that this is the same noise level as for the subtraction of two signals with uncorrelated noise sources.

\section{Phase noise}
Phase noise on the input signal presents the second major source for fluctuations of the output phase. In the following we shall derive an analytical expression for the output phase fluctuations after two input signals affected by phase noise have been processed by the phase-meter and their phase difference was found. We make again the following ansatz to describe the phase-meter processing of one input signal
\begin{eqnarray}
\Re(F_t)&=&\sum_{k=0}^{N-1}\left[A_{\rm in} \cos\left(\frac{2\pi m k }{N}-\phi-n_k\right)\right]\cos\left(\frac{2\pi m k }{N}\right)\nonumber\\
\Im(F_t)&=&\sum_{k=0}^{N-1}\left[A_{\rm in} \cos\left(\frac{2\pi m k }{N}-\phi-n_k\right)\right]\sin\left(\frac{2\pi m k }{N}\right),\nonumber\\ 
\end{eqnarray}
where $n_k$ denotes the phase fluctuations of the input phase at each sample step $k$. As in the case of amplitude noise, the phase fluctuations are assumed to be uncorrelated with properties described by Eq.\ref{equ:noise_operators}.
As phase noise is contained within the argument of the input signal, it is more difficult to treat mathematically than amplitude noise and the derivation is somewhat lengthy so that we restrict ourselves to an overview over the major steps.
The input signal arguments are expanded according to basic trigonometric relations and sine and cosine terms containing the phase noise (pn) fluctuations are approximated by the following relations
\begin{eqnarray}
\cos(\phi+n_k)&=&\cos\phi-n_k\sin\phi\nonumber\\
\sin(\phi+n_k)&=&\sin\phi+n_k\cos\phi\nonumber\\
\label{equ:sine_expansion}
\end{eqnarray}
We then find after some algebra
\begin{equation}
\tan\phi_{\rm pn}=\frac{\Re(F_t)}{\Im(F_t)}=\frac{\sin\phi(1-\alpha)+\beta\cos\phi}{\cos\phi(1+\alpha)-\gamma\sin\phi},
\label{equ:tan_phi}
\end{equation}
where we introduced the following abbreviations
\begin{eqnarray}
\alpha&=&\frac{2}{N}\sum_{k=0}^{N-1}\cos\left(\frac{2\pi m k}{N}\right)\sin\left(\frac{2\pi m k}{N}\right)n_k\nonumber\\
\beta&=&\frac{2}{N}\sum_{k=0}^{N-1}\sin^2\left(\frac{2\pi m k}{N}\right)n_k\nonumber\\
\gamma&=&\frac{2}{N}\sum_{k=0}^{N-1}\cos^2\left(\frac{2\pi m k}{N}\right)n_k\nonumber\\
\label{equ:PN_definitions}
\end{eqnarray}
Equation \ref{equ:tan_phi} describes the phase-meter output phase for a signal with given phase noise and is used in the next step to calculate the common mode noise rejection between two signals.
\subsection{Common Mode Phase Noise Rejection}
We are interested in finding the noise rejection in the phase difference between two signals, one of center frequency $\phi$ and the other of center frequency $\psi$ relative to the phase-meter reference clock.  Both signals are affected by the same phase fluctuations and we aim to determine the noise rejection from the differential phase $\phi_{\rm pn}-\psi_{\rm pn}$. The following trigonometric identity is useful for this purpose
\begin{equation}
\tan(\phi-\psi)=\frac{1}{\cot\phi+\tan\psi}-\frac{1}{\tan\phi+\cot\psi}
\label{equ:PN_trig_property}
\end{equation}
Inserting the definition of Eq.\ref{equ:tan_phi} for $\tan(\phi_{\rm pn})$ and an analogous one for $\tan(\psi_{\rm pn})$ into Eq.\ref{equ:PN_trig_property}, and neglecting terms to second order or higher in the noise fluctuations $n_k$, we find after some lengthy algebra
\begin{eqnarray}
&&\tan(\phi_{\rm pn}-\psi_{\rm pn})=\nonumber\\
&&\frac{\sin(\phi-\psi)}{\cos(\phi-\psi)+2\alpha\cos(\phi+\psi)+(\beta-\gamma)\sin(\phi+\psi)}\nonumber\\
\label{equ:PN_rejection}
\end{eqnarray}
Defining abbreviations for the phase difference $x=\phi-\psi$ and the sum of all the small noise quantities $\epsilon=2\alpha\cos(\phi+\psi)+(\beta-\gamma)\sin(\phi+\psi)$ we may rewrite Eq.\ref{equ:PN_rejection} as follows:
\begin{eqnarray}
f(x,\epsilon)=\phi_{\rm pn}-\psi_{\rm pn}&=&\arctan\left(\frac{\sin x}{\cos x+\epsilon}\right)\nonumber\\
\approx x-\epsilon\sin x,
\label{equ:PN_approx}
\end{eqnarray}
where for the last step we performed a Taylor expansion to first order in $\epsilon$ of $f(x,\epsilon)$. In conclusion, for the measured phase fluctuations which are induced by the phase noise affecting both signals common mode we find after some simplifications:
\begin{eqnarray}
\Delta n_{\rm pn}&=&(\phi_{\rm pn}-\psi_{\rm pn})-(\phi-\psi)=-\epsilon\sin(\phi-\psi)\nonumber\\
=&-&\frac{2}{N}\sin(\phi-\psi)\cos(\phi+\psi)\sum_{k=0}^{N-1}\sin\left(\frac{4\pi m k}{N}\right)n_k\nonumber\\
&+&\frac{2}{N}\sin(\phi-\psi)\sin(\phi+\psi)\sum_{k=0}^{N-1}\cos\left(\frac{4\pi m k}{N}\right)n_k\nonumber\\
\end{eqnarray}
It is interesting to note that the fluctuations are only fully suppressed if the differential phase between the two input signals is exactly zero, i.e. $\phi-\psi=0$, whilst the fluctuations are maximal at an amplitude of $\epsilon$ when the two signals are out of phase by $\pi/2$.
In the next step we determine the mean-square value of the noise fluctuations and obtain after discarding the cross-term
\begin{eqnarray}
&&\langle \Delta n_{\rm pn}^2\rangle=\sin^2(\phi-\psi)\frac{4}{N^2}\cos^2(\phi+\psi)\sum_{k=0}^{N-1}\sin ^2 \left(\frac{4\pi m k}{N}\right)\langle n_k^2\rangle\nonumber\\
&&+\sin^2(\phi-\psi)\frac{4}{N^2}\sin^2(\phi+\psi)\sum_{k=0}^{N-1}\cos ^2 \left(\frac{4\pi m k}{N}\right)\langle n_k^2\rangle\nonumber\\
&&=\frac{2}{N}\langle n_k^2\rangle\sin^2(\phi-\psi)
\label{equ:PN_result}
\end{eqnarray}
Defining an input phase noise of uniform linear spectral density $n_{\rm ld}$ in units of ${\rm rad}/\sqrt{Hz}]$ we obtain from Eq.\ref{equ:PN_result}:
\begin{eqnarray}
\sqrt{\langle \Delta n_{\rm pn}^2\rangle}&=& n_{\rm ld}\sqrt{f_{\rm up}}\left|\sin(\phi-\psi)\right|\label{equ:PN_lsd}\\
{\rm LSD}(\Delta n_{\rm pn})&=& n_{\rm ld}\sqrt{2}\left|\sin(\phi-\psi)\right|
\label{equ:PN_main_result}
\end{eqnarray}

\begin{figure}
   \centerline{\includegraphics[width=.8\columnwidth]{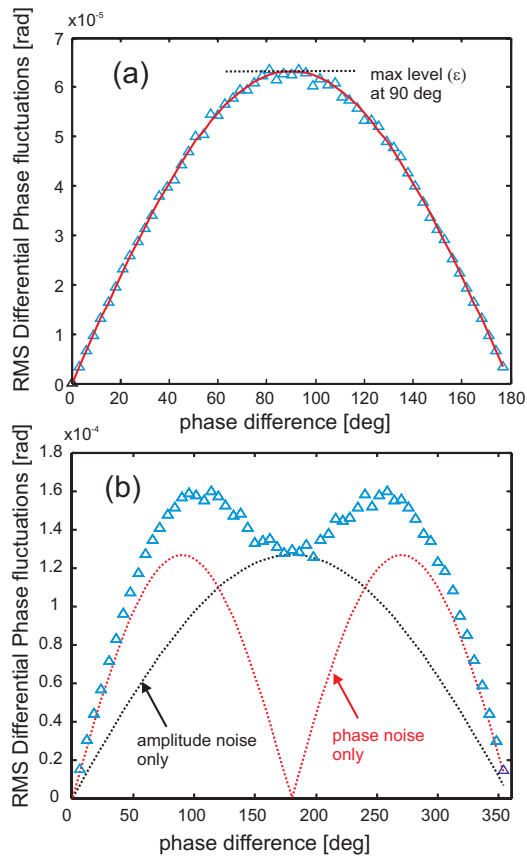}}
   \caption
    { \label{fig:phase_noise_plot}
(Color online) The differential output noise $\Delta n$ is plotted against the mean phase difference $\phi-\psi$ between signal 1 and signal 2: (a) Phase noise only is applied to the input signal with RMS values at each sampling step of $10^{-3}{\rm rad}$. The predicted curve (Eq.\ref{equ:PN_lsd}) is given by the red solid line, the results of Monte Carlo simulations are given by the blue triangles. (b) Phase noise of RMS $2\times 10^{-3}{\rm rad}$ and amplitude noise of RMS $10^{-3}$ are applied. The open triangles denote the exact results from Monte Carlo simulations. The predicted curves if either of the noise sources is applied alone are given by the red and black dotted lines, respectively.}
\end{figure}
Equation \ref{equ:PN_main_result} is one major result of this paper. The analytical results have been verified by comparison against Monte Carlo simulations, which is depicted in Fig.\ref{fig:phase_noise_plot}a. Excellent agreement is found between the analytical prediction of Eq.\ref{equ:PN_lsd} (red solid line) and the output data from the simulation (blue triangles), where the standard deviation of 2000 simulation runs per angle are plotted.
The output phase noise increases according to a sine law and reaches a maximum at 90 deg phase difference between the two signals. \\
\section{Comparison to experimental observations}
The situation becomes more complex if both, amplitude and phase noise, affect the input signals, which is depicted in Fig.\ref{fig:phase_noise_plot}b. In this case, the resulting differential output noise follows neither the prediction for phase noise given by Eq.\ref{equ:PN_lsd} nor the one for amplitude noise given by Eq.\ref{equ:AN_key_equation} but exhibits certain characteristic features of both. If either source dominates but one does not know which one, it is easy to distinguish between the two experimentally due to the different extrema and periodicity of the output noise as a function of $\phi-\psi$: The impact of phase noise is minimal for $\phi-\psi=\pi$ and has a period of $\pi$, whereas the impact of amplitude noise is maximal at $\phi-\psi=\pi$ and has a period of $2\pi$.\\
During recent test campaigns of the LPF optical metrology system seemingly random changes of the measurement noise floor (in certain spectral regions) were observed in between measurements, which upon closer inspection seemed to correlate with changes in the relative phase $\phi-\psi$ between measurement and reference interferometer (see e.g. \cite{Hec2011}). As the relative phase depends on the test-mirror position which is variable, this leads to a correlation between the measured noise floor and the mirror position. These observations are in agreement with the predictions made in this paper for the effect of high frequency phase noise but more systematic measurements would have to be performed to clearly confirm such a relationship. It should be pointed out that frequency mixing and nonlinearities due to imperfect separation of polarization states may lead to periodic errors in polarizing heterodyne interferometers \cite{Bob1999}. However, the optical metrology system of LPF is non-polarizing and such errors can therefore be ruled out as a possible explanation of the measured effects.\\
Interestingly, it has also been pointed out that relations Eq.\ref{equ:PN_lsd} and Eq.\ref{equ:AN_key_equation} are similar to equations previously obtained to describe the impact of optical sidebands spaced around the center frequency at multiples of the heterodyne frequency\cite{Wan2006}. The detrimental effects of those were overcome by inclusion of an OPD actuator for phase stabilization, as depicted in the system schematic of Fig.\ref{fig:opt_bench}.
\section{Conclusion}
We have derived expressions describing the coupling of high frequency amplitude and phase noise of an input signal into phase fluctuations of the digital phase-meter output of a heterodyne interferometer. In a next step we calculated the expected common mode noise suppression between two input signals and found a dependency on $\sin(\phi-\psi)$ for phase noise and on $\sin[(\phi-\psi)/2]$ for amplitude noise, where $\phi-\psi$ describes the phase difference between the two input signals.
All analytical derivations were complemented by numerical Monte Carlo simulations which were found to be in excellent agreement and to validate any approximations which were made. Recent experimental observations of correlations between measurement phase and noise floor in the optical metrology system of the LISA Pathfinder mission could possibly be ascribed to the effects discussed in this paper.
\acknowledgments     
The author would like to thank David Hoyland (University of Birmingham) and Vinzenz Wand (now OHB System) for their help, useful information and stimulating discussions.
He also gratefully acknowledges useful discussions with Reinhold Flatscher, Nico Brandt, Patrick Bergner, Tobias Ziegler, R\"udiger Gerndt and Ulrich Johann (EADS Astrium), as well as Gerhard Heinzel (Albert Einstein Institute) and Paul McNamara (European Space Agency).


\end{document}